\documentclass[aps,prl,twocolumn,amsmath,floatfix,citeautoscript]{revtex4}
\usepackage{graphicx}
\usepackage{dcolumn}
\usepackage{multirow}
\usepackage{array}
\usepackage{tabularx}

\begin{document}


\title{Stabilization of highly polar BiFeO$_3$-like structure: a new interface design route for enhanced
ferroelectricity in artificial perovskite superlattices}
\author{Hongwei Wang$^{1}$}
\author{Jianguo Wen$^{2,*}$, Dean J. Miller$^{2}$}
\author{Qibin Zhou$^{4}$}
\author{Mohan Chen$^{5}$}
\author{Ho Nyung Lee$^{3}$}
\author{Karin M. Rabe$^{4}$}
\author{Xifan Wu$^{1,*}$}
\affiliation{$^{1}$Department of Physics, Temple University, Philadelphia, PA 19122, USA }
\affiliation{$^{2}$Center for Nanoscale Materials, Argonne National Laboratory, Argonne, IL 60439, USA}
\affiliation{$^{3}$ Materials Science and Technology Division, Oak Ridge National Laboratory, Oak Ridge, Tennessee 37831, USA}
\affiliation{$^4$Department of Physics and Astronomy, Rutgers University,
Piscataway, NJ 08854-8019, USA}
\affiliation{$^{5}$ Department of Mechanical and Aerospace Engineering, Princeton University, Princeton, New Jersey 08544, USA}

\date{\today}
\begin{abstract}
In ABO$_3$ perovskites, oxygen octahedron rotations are common structural distortions that 
can promote large ferroelectricity in BiFeO$_3$ with an $R3c$ structure~\cite{NeatonBFO}, but suppress
ferroelectricity in CaTiO$_3$ with a $Pbnm$ symmetry~\cite{Benedek_JPCC}.  
For many CaTiO$_3$-like perovskites,
the BiFeO$_3$ structure is a metastable phase. Here, we report the stabilization of
the highly-polar BiFeO$_3$-like phase of CaTiO$_3$ in a BaTiO$_3$/CaTiO$_3$
superlattice grown on a SrTiO$_3$ substrate.
The stabilization is realized by a reconstruction of oxygen 
octahedron rotations at the interface from the pattern of nonpolar bulk CaTiO$_3$ to a 
different pattern that is characteristic of a BiFeO$_3$ phase. 
The reconstruction is interpreted through a combination of amplitude-contrast
sub $0.1 nm$ high-resolution transmission electron microscopy and first-principles
theories of the structure, energetics, and polarization
of the superlattice and its constituents. 
We further predict a number of new artificial ferroelectric materials 
demonstrating that nonpolar perovskites can be turned into
ferroelectrics via this interface mechanism.
Therefore, a large number of perovskites with the CaTiO$_3$ structure type,
which include many magnetic representatives, are now good candidates as novel highly-polar
multiferroic materials~\cite{{Physics_Today}}.

\end{abstract}


\maketitle


\def\scr{\scriptsize}

\section{Introduction}
New mechanisms to generate ferroelectricity (FE) have recently been the subject of active research,
due to both fundamental interest and the technological importance
of ferroelectrics and related materials~\cite{Dawber_Review}.
Novel ferroelectrics have potentially higher performance
for practical applications, as well as potential compatibility with other functional properties such
as magnetism, yielding multiferroics and other multifunctional
materials~\cite{Physics_Today,Schlom_Nature_2010,Benedek_hybrid_improper}.

Artificially structured perovskite superlattices offer rich opportunities for novel
ferroelectricity~\cite{Rabe_Science, Lee_Nature_tricolor,Warusawithana_tricolor_PRL,Sai_PRL_2000,Bousquet_Nature}.
Non-bulk phases for the constituent layers can be stabilized
by the mechanical and electrical boundary conditions characteristic of a
superlattice~\cite{Rabe_Review_2007,Stengel_Nature_Physics},
potentially turning constituents that are nonpolar in bulk form into ferroelectrics~\cite{Neaton_2003, Eom_STO_2010}.
Competing low-energy metastable phases can be readily found in perovskites with low tolerance
factors, promoting oxygen octahedron rotation (OOR) instabilities along the Brillouin-zone-boundary
R-M line. The ground state structure in such cases is generally the nonpolar orthorhombic $Pbnm$
structure. As a typical example, the oxygen octahedron in a CaTiO$_3$ (CTO) can be described by
its rotation around [110] axis and an in-phase rotation around [001] axis (a$^-$a$^-$c$^+$ in Glazer notation).
Such a pattern of  OOR favors antipolar behavior instead of FE~\cite{Benedek_JPCC}.
On the other hand, OOR with a different pattern can also promote large FE.
As one famous example, in BiFeO$_3$ (BFO) with $R3c$ structure,
the oxygen octahedron can be characterized by a rotation around [110] and an out-of-phase
rotation around [001], yielding a fairly large polarization along [111] (a$^-$a$^-$a$^-$ in Glazer notation).
Compared to the widespread of CTO-like materials,  BFO-like perovskites are relatively rarely seen.
As a result, the OOR is generally thought to suppress FE in perovskites.

\begin{figure}
\includegraphics[width=3.35in]{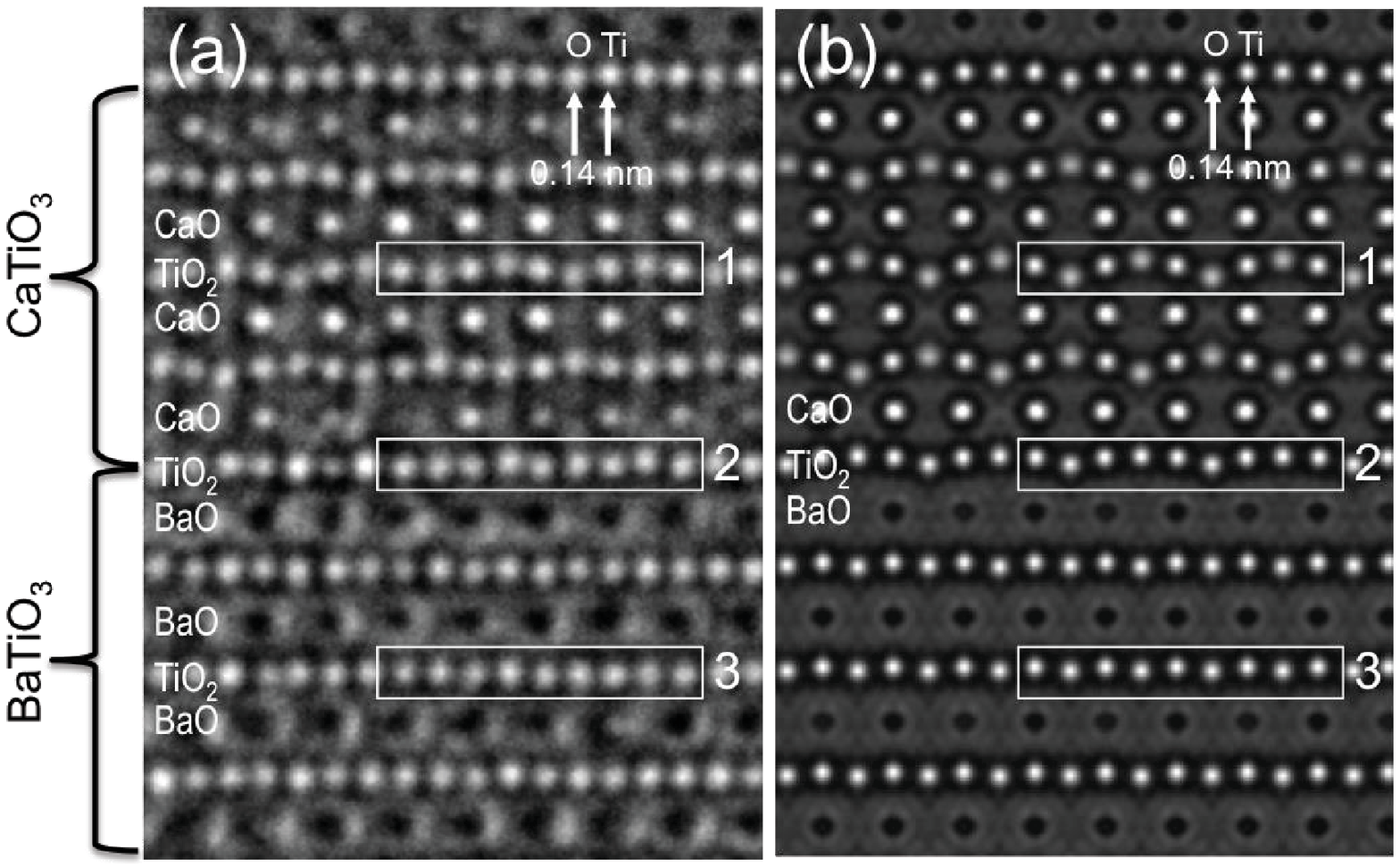}
\caption{\label{fig:TEM}
{\bf Experimental and simulated HRTEM images showing oxygen octahedral tilts in the 4BTO/4CTO superlatice film
grown on a STO substrate along [110].} (a). Experimental HRTEM image using an amplitude contrast imaging
method ($ Cs = 3\mu m,Cc=1\mu m, \Delta f=-1nm $). BaO columns (dark dots) and CaO columns (bright dots) show different
channeling contrast. Oxygen atomic columns displace differently, either upward or downward,
with respect to the central Ti atoms in box 1, 2 and 3. 
$0.14 nm$ indicates the spacing between Ti and O columns. Accumulated electron dose is $3 \times 10^4$ electrons/$nm^2$. (b). Simulated HRTEM image using the atomic positions
obtained from first-principles calculations. The simulated image matches well with the experimental 
image except a sharper column contrast in the simulated image compared to the experiment one because 
electron beam induced object excitations are omitted as pointed out by 
Kisielowski et al. (see Supplementary Materials S4) ~\cite{Kisielowski}. }
\end{figure}
\begin{figure*}
\includegraphics[width=7.0in]{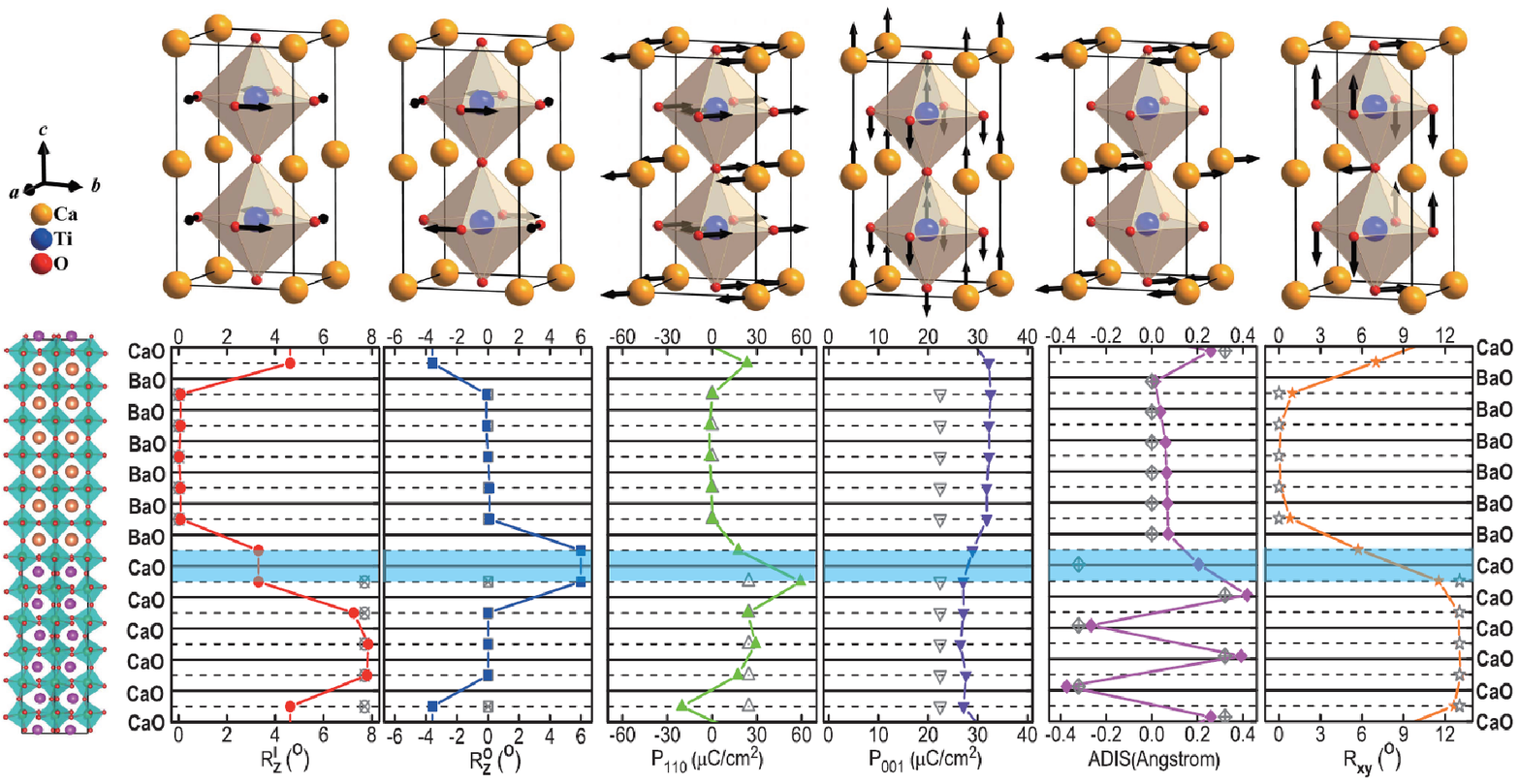}
\caption{\label{fig:superlattices}
{\bf Computed local properties associated with ferroelectric and non-polar modes in 6BTO/6CTO superlattice and 
schematic plots for the atomic displacements of the oxygen octahedron rotation and ferroelectric modes.}
First-principles calculations of 6BTO/6CTO superlattice showing layer-by-layer decompositions
in-phase oxygen octahedron rotation around [001] $\rm R_z^i$, out-of-phase oxygen octahedron rotation around [001]
$\rm R_z^o$, in-plane polarization P$_{110}$, out-of-plane polarization P$_{001}$,
antipolar modes (AFE) represented by A-site cation displacements (ADIS), and oxygen octahedron rotation around [110] 
$\rm R_{xy}$. The corresponding strained bulk values are also denoted by the open symbol in the plot, and in particular, the 
open strained bulk values in P$_{001}$ are predicted by the {\it dielectric slab} model~\cite{Neaton_2003}.
A highly polar BiFeO$_3$-like interface phase in CTO is highlighted in blue.
}
\end{figure*}

However, for many perovskites, the BFO-like structure serves as a low-energy metastable phase~\cite{Benedek_JPCC}.
Therefore, it would be beneficial if an artificial perovskite superlattice could 
stabilize this metastable phase for the entire constituent layers or in a region near interface.
To this end, a reliable design mechanism can be derived only from precisely determined atomic positions
in experiments followed by theoretical interpretations based on first-principles calculations.

\section{Experimental and first-principles results}
Aberration-corrected high-resolution transmission electron microscopy (HRTEM) is a powerful method for accurate
visualization of oxygen octahedron distortions~\cite{Jia_TEM_2003,Jia_TEM_2008}.
Recently, it was shown that amplitude contrast imaging in HRTEM could be used to discriminate
heavy and light element columns based on channeling contrast~\cite{ACI}, allowing one to locate the exact
interface and to visualize OOR angles in different atomic layers (see Supplementary Materials S1).
Fig.~\ref{fig:TEM}(a) shows an experimental HRTEM image of
a 4BaTiO$_3$(BTO)/4CTO superlattice film along the [110] direction
of the SrTiO$_3$ (STO) substrate. This image was obtained by correcting both spherical and chromatic aberrations
to achieve amplitude contrast imaging conditions (Cs = 3 $\mu m$, Cc = $1\mu m$).
In this image, channeling contrast between Ca and Ba columns is clearly observed: atomic columns of CaO and BaO
appear as bright and dark dots, respectively; oxygen and Ti columns appear as bright dots.
Due to the interdiffusion of Ba and Ca at the interface, the intensity at A site varies depending 
on the ratio of Ca and Ba as discussed in detail in the supplementary material (S2). 
It is seen that BTO and CTO grow coherently on the STO substrate, showing the same
in-plane lattice constant as that of STO, and elongated c-axis in the BTO layer and shortened c-axis in the CTO layer
(see Supplementary Materials S3).
Within the CTO layer of the superlattice (box 1),
a strongly corrugated TiO$_2$ plane is observed in which the oxygen atoms displace upward and downward with
respect to the central Ti atoms, corresponding to an OOR around [110] by 9$^{\circ}$, comparable to that of bulk CTO.
For TiO$_2$ planes between two BaO planes (box 3), alternating displacement of the oxygen atoms,
and thus the amplitude of the OOR, is negligible, consistent with the fact that bulk BTO strongly resists
OORs. For TiO$_2$ planes between one BaO and one CaO plane (box 2),
the OOR around [110] is 3$^{\circ}$, smaller than that in the interior of the CTO layers.

For comparison, in Fig.~\ref{fig:TEM}(b) we present the simulated HRTEM image using the
atomic positions of the 4BTO/4CTO superlattice obtained from first-principles calculations.
The simulated HRTEM image for the computed structure shows the same pattern of 
OOR as in the experiment (compare boxes 1, 2, and 3
in Fig.~\ref{fig:TEM}(a) and (b)), with amplitudes of 12.5$^{\circ}$ in the CTO layer and 5.5$^{\circ}$ at the interface.
The quantitative difference in OOR around [110] angles from the experimental observation can be partly attributed to the fact that
the experiments were performed at T= 300 K, while the ground state structural relaxation by density functional theory was at T = 0 K.
In addition, in this image it is possible to discern the small uniform
displacement of the oxygens relative to the Ti atoms in the TiO$_2$ plane, which is associated with the
spontaneous polarization of the superlattice. While this
displacement is present in all the TiO$_2$ layers, it can be more easily identified in those belonging to
the interior BTO layers, which do not have the corrugation associated with OOR.

We use the atomic-scale information from the first-principles results for a 
detailed layer-by-layer investigation of the properties of the superlattice.
We focus our discussion on the 6BTO/6CTO superlattice, which allows a clearer distinction between the interface and interior layers;
the corresponding results for the 4BTO/4CTO superlattice are similar (see Supplementary Materials S6).
The computed spontaneous polarization is 29 $\mu C cm^{-2}$ along [001] and 11 $\mu C cm^{-2}$ along [110].
The resulting layer-by-layer decomposed structural distortions and polarizations are shown in Fig.~\ref{fig:superlattices}.

\begin{table}[ht]
\begin{center}
\caption{Computed bulk properties of 
 $\rm R_z^i(\circ)$, $\rm R_z^o(\circ)$, $\rm R_{xy}(\circ)$, ADIS ($\AA$), P$_{001}$($~\mu C/cm^2$), P$_{110}$($~\mu C/cm^2$), 
total polarization $P_{\rm T}$($~\mu C/cm^2$), $c/a$ ratio, and the total energies ($eV$) 
for strained BTO and CTO on STO substrate modeled in 20-atom supercells. 
Both fixed electric ($E$) field and displacement ($D$) field boundary conditions are considered, 
which are used to described the electric boundary conditions of a perovskite in
its nature bulk or within an insulating superlattice. }
\begin{tabularx}{0.46\textwidth}{ l | c c c | c c }
\cline{1-6}
  \hline
   \hline
{\footnotesize Boundary}  & \multicolumn{3}{c}{  Fixed $E$ field} &\multicolumn{2}{|c}{ Fixed $D$ field} \\
{\footnotesize Condition} & \multicolumn{3}{c}{$E=0~V/m$} &\multicolumn{2}{|c}{$D= 29~\mu C/cm^2$} \\
  \hline
Perovskite & BaTiO$_3$ & CaTiO$_3$ & CaTiO$_3$ & CaTiO$_3$ & CaTiO$_3$  \\
 \hline
{\footnotesize Symmetry}  & P4mm & Pbnm & e-R3c & Pbnm & e-R3c  \\
 R$_{z}^{i}$ & 0 & 7.7 & 0 & 7.6 & 0\\
 R$_{z}^{o}$ & 0 & 0 & 7.3 & 0 & 7.2 \\
 R$_{xy}$ & 0 & 12.7 & 11.5 & 12.8 & 11.7  \\
 ADIS & 0 & 0.29 & 0 & 0.32 & 0 \\
 $P_{001}$ & 41.93 & 0 & 20.01 & 27.81 & 28.96 \\
 $P_{110}$ & 0 & 26.40 & 52.19 & 24.53 & 51.25 \\
 $P_{\rm T}$ & 41.93 & 26.40 & 55.89 & 37.08 & 58.87 \\
 $c/a$ & 1.067  & 0.964 &0.968 & 0.965 & 0.967\\
 Energy & -162.645 & -164.158 & -164.066 & -164.106 & -164.040\\
   \hline
  \hline
\end{tabularx}
\label{tab:bulk_value}
\end{center}
\end{table}

According to the {\it dielectric slab} model~\cite{Neaton_2003}, the structure of the constituent layers of the superlattice
should be closely related to those of strained bulk materials under the electrical boundary condition of a fixed 
displacement ($D$) field, imposed by the superlattice as summarized in Table~\ref{tab:bulk_value}. 
Indeed, as shown in Fig.~\ref{fig:superlattices},
the interior BTO layers have negligible OOR with a polarization of 32 $\mu C/ cm^{2}$ along the [001] direction.
This is consistent with the structure and large polarization of strained BTO; the reduction from
the strained bulk value of 42 $\mu C/ cm^{2}$
can be attributed to the electrostatic cost of polarizing the nonpolar CTO layer.
Both bulk CTO and strained bulk CTO are characterized by the strong OORs due to structural instabilities at the
zone-boundary $M$ and $R$ points.
Therefore, the interior CTO layers are dominated by $R_{xy}$ and $R^{i}_{z}$, which are
OOR around [110] and an in-phase OOR around [001] respectively as shown in Fig.~\ref{fig:superlattices}.
In addition, a large antipolar (AFE) mode develops in the CTO layers that
can be clearly identified by the zig-zag movement of A-site displacement along [110] direction.
It should be stressed that this antipolar distortion is a structural distortion at the $X$ point favored
by the trilinear coupling due to the pattern of OOR in 
CTO-like materials. The above distortion in the interior CTO layers can be
clearly seen in Fig.~\ref{fig:superlattices}, as well as
in the TEM image in Fig.~\ref{fig:TEM} (a) (see Supplementary Materials S5).
This AFE mode was also recently pointed out to be the key to the 
suppressed FE in all CTO-like perovskites~\cite{Benedek_JPCC}.
Due to the applied tensile epitaxial strain, the interior CTO layers are
polar along [110] direction with a magnitude of $26.4~\mu c/cm^2$ just like the strained CTO~\cite{Eklundpaper2009}.

If the interface effect is negligible, the {\it dielectric slab} model can be used to predict the polarization,
yielding a value of $22.4~ \mu C/ cm^{2}$
along the [001] direction. The first-principles calculation gives $P_{001} =29.0~ \mu C/ cm^{2}$.
The discrepancy from the {\it dielectric slab} model suggests that the interface effect cannot be neglected.
Such a large enhancement of the polarization ($\sim 25\%$) is a strong indication of a highly polar
interface reconstruction.
Indeed, examination of Fig.~\ref{fig:superlattices} reveals that the structure 
at the interface of the CTO layers differs significantly
from that of the strained bulk CTO, with the OOR being suppressed at the interface of the superlattice.
The AFE type displacement, which is driven by the trilinear coupling~\cite{Benedek_JPCC}
involving the OORs of CTO, is suppressed too. Furthermore, a new structural pattern of OOR emerges at the interface:
an OOR around the [110] axis and an out-of-phase OOR around the [001] axis
for a TiO$_6$ sandwiched between two interface CaO layers, with rotation angles comparable to those of the
strained bulk in-phase OOR. This new structure pattern is exactly the same as one would
observe for oxygen octahedron rotation in BiFeO$_3$ and similar perovskites with $R3c$ symmetry.

\section{Microscopic mechanism}

Here, we propose that this change in structure
at the interface can be interpreted as the local stabilization of a BFO-like structure
different from that of the bulk CTO. As far as the topology of the oxygen octahedron
rotation network is concerned, oxygen octahedra in both BFO and CTO
rotate around [110]; however, BFO differentiates itself from CTO by its out-of-phase
OOR around [001] instead of the in-phase counterpart in CTO.
The out-of-phase and in-phase OOR around [001] originate from symmetry-nonequivalent
structural instabilities at the $R$ and $M$ points respectively.
This stabilization of a BFO-like structure in CTO layers near the interface
is derived from the metastable polar e-$R3c$ phase and is
compatible with a much larger polarization than 
that in bulk CTO as shown in Table~\ref{tab:bulk_value}.
It has been shown that this phase cannot be stabilized relative to the $Pbnm$
phase by epitaxial strain alone~\cite{Eklund_Thesis}. However, in the superlattice,
the suppression of the tilt angles by proximity to BTO, assisted by the electrical and mechanical boundary
conditions that favor a phase with a component of polarization along [001], is sufficient
to stabilize the structure~\cite{Zhou_Thesis}.

To explore the stabilization of this phase more quantitatively, we constructed first-principles based
models for the strained $Pbnm$ phase (designated $E_{\rm ``CTO''}({\rm R_z^i},{\rm R}_{\rm xy},
{\rm \rm AFE_{xy}},{\rm FE_{xy}})$) and for the metastable
e-$R3c$ phase ($E_{\rm ``BFO''}({\rm R_z^o},{\rm R}_{\rm xy},{\rm FE_{xy}},{\rm FE_z})$).
Facilitated by space group symmetry analysis, the models of both
$E_{\rm ``CTO''}$ and $E_{\rm ``BFO''}$ are built through polynomial
expansions of the total energy from first-principles calculations with respect to the high-symmetry reference structure 
($P4/mmm$ phase) in terms of the amplitudes of the relevant modes.
In the above, ${\rm R_z^i}$, ${\rm R_z^o}$, ${\rm R}_{\rm xy}$, ${\rm AFE_{xy}}$, ${\rm FE_{xy}}$, ${\rm FE_z}$
represent the mode amplitude of  in-phase OOR around [001], out-of-phase OOR around [001], OOR around [110],
in-plane antipolar mode and in-plane and out-of-plane FE modes respectively. The resulting models are shown in the following for
$E_{\rm ``CTO''}$ and $E_{\rm ``BFO''}$ respectively as (see Supplementary Materials S7 for fitted coefficients):

\begin{equation}
\begin{array}{r@{~}l}
E_{\tiny \rm ``CTO''} & =  a_{1}{\rm R^i_z}^{2}+b_{1}{\rm R^i_z}^{4}+a_{2}{\rm R}_{\rm xy}^{2}
+b_{2}{\rm R}_{\rm xy}^{4}+a_{3}{\rm AFE_{xy}^{2}}\cr
&+b_{3}{\rm AFE_{xy}^{4}} + a_{4}{\rm FE_{\rm xy}^{2}}+b_{4}{\rm FE_{\rm xy}^{4}}
+ c_{1}{\rm R^i_z}^{2}{\rm R}_{\rm xy}^{2} \cr
& +c_{2}{\rm R^i_z}^{2}{\rm AFE_{xy}^{2}} +\ c_{3}{\rm R^i_z}^{2}{\rm FE_{\rm xy}^{2}}
+ c_{4}{\rm R}_{\rm xy}^{2}{\rm AFE_{xy}^{2}} \cr
& +c_{5}{\rm R_{xy}^2}{\rm FE_{xy}^{2}}+c_{6}{\rm AFE_{xy}^{2}}{\rm FE_{\rm xy}^{2}}\cr
&+ \ d_{1}{\rm R^i_z}{\rm R}_{\rm xy}{\rm AFE_{xy}},
\label{eq:LP}
\end{array}
\end{equation}
\begin{equation}
\begin{array}{r@{~}l}
E_{\rm ``BFO''}&= \alpha_{1}{\rm R^o_z}^{2}+\beta_{1}{\rm R^o_z}^{4}+\alpha_{2}{\rm R}_{\rm xy}^{2}
+\beta_{2}{\rm R}_{\rm xy}^{4}+\alpha_{3}{\rm FE_{z}^{2}} \\
 &+\ \beta_{3}{\rm FE_{z}^{4}}+\alpha_{4}{\rm FE_{\rm xy}^{2}} +  \beta_{4}{\rm FE_{\rm xy}^{4}}
+\gamma_{1}{\rm R^o_z}^{2}{\rm R}_{\rm xy}^{2} \cr
& + \gamma_{2}{\rm R^o_z}^{2}{\rm FE_z^{2}} +\ \gamma_{3}{\rm R^o_z}^{2}{\rm FE_{xy}^{2}}
+\gamma_{4}{\rm R}_{\rm xy}^{2}{\rm FE_z^{2}} + \cr
& \gamma_{5}{\rm R}_{\rm xy}^{2}{\rm FE_{xy}^{2}}
+\gamma_{6}{\rm FE_z^{2}}{\rm FE_{xy}^{2}} +\kappa_{1}{\rm R^o_z}{\rm R}_{\rm xy}{\rm FE_z FE_{xy}.}
\label{eq:HP}
\end{array}
\end{equation}
\begin{figure}
\includegraphics[width=3.4in]{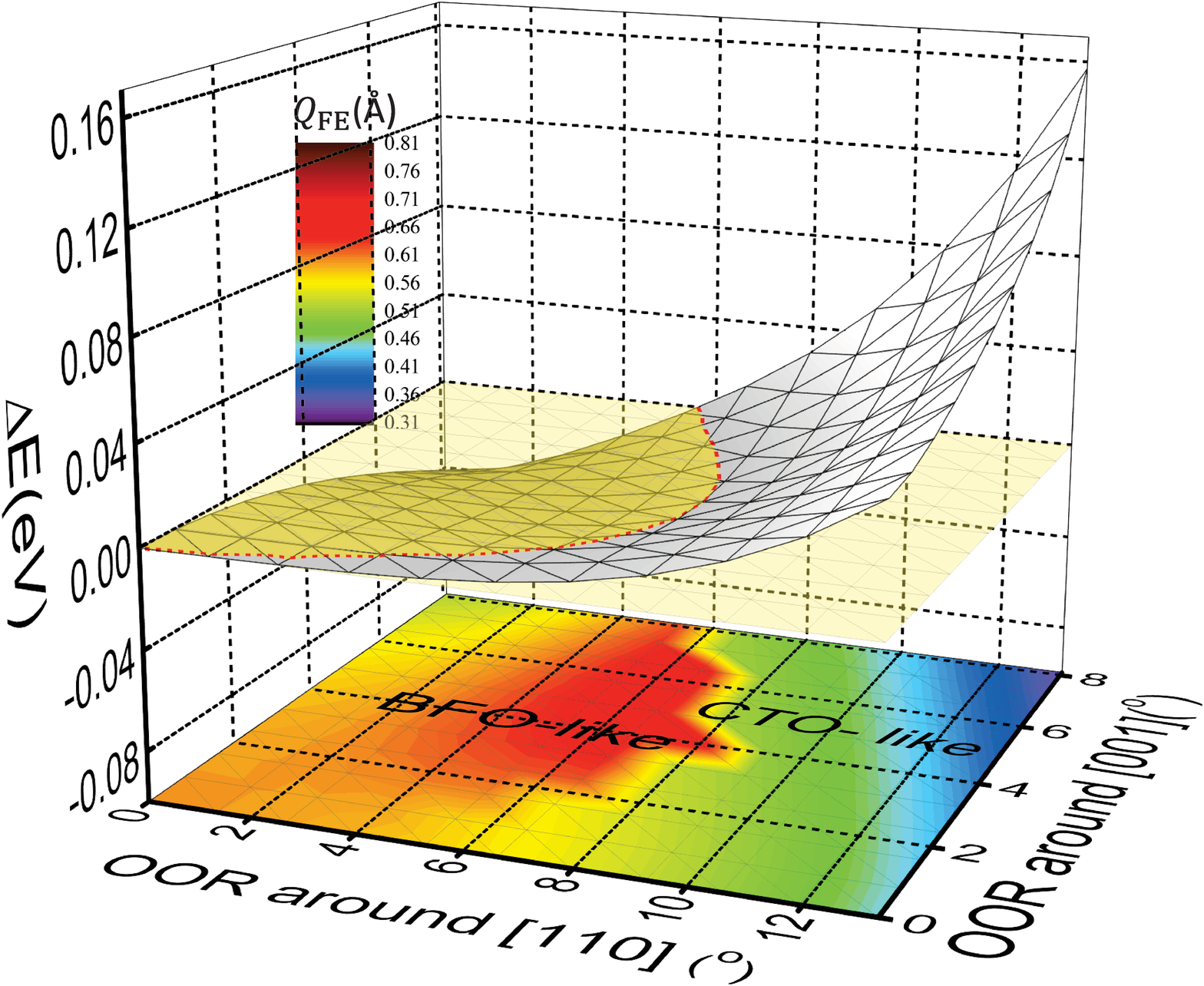}
\caption{\label{fig:phase_transition}
{\bf Appearances of both BiFeO$_3$-like  and CaTiO$_3$-like 
phases and the transition from one to the other as functions of oxygen octahedral 
rotations in strained bulk CaTiO$_3$.}
Phase stabilities studied by the relative energetics between BFO-like and CTO-like phase as 
plotted by $\Delta E$ as functions of magnitudes of oxygen octahedral rotation around [001] and [110] 
respectively. The FE mode amplitudes are represented by projected colors in the base as well.}
\end{figure}
Assuming the angles of the oxygen octahedron rotations are tunable parameters 
under experimental conditions, we further define the functions
$\displaystyle \mathcal{F}_{\rm ``CTO''}({\rm \rm R^i_z},{\rm R}_{\rm xy})=
\min_{\rm AFE_{xy},  FE_{\rm xy}}E_{\rm ``CTO''}({\rm R^i_z},{\rm R}_{\rm xy},{\rm AFE_{xy}}, {\rm FE_{\rm xy})}$,
and $\displaystyle\mathcal{F}_{\rm ``BFO''}({\rm R_z^o},{\rm R}_{\rm xy})=
\min_{{\rm FE_{xy}},{\rm FE_z}}E_{\rm ``BFO''}({\rm R_z^o},{\rm R}_{\rm xy},{\rm FE_{xy}},{\rm FE_z})$.
In order to understand how the BFO-like phase can be stabilized relative to the CTO-like phase,
we then evaluate $\Delta E= \mathcal{F}_{\rm ``BFO''}-\mathcal{F}_{\rm ``CTO''}$ as functions of oxygen octahedron
rotation magnitudes around [110] and [001].
The resulting $\Delta E$ is presented in Fig.~\ref{fig:phase_transition}. The total FE mode amplitudes are also
presented by the color spectrum in the base plane in Fig.~\ref{fig:phase_transition}.
It can be seen that when the angles are fixed to the values of bulk CTO regions in the superlattice, as shown 
in Fig.~\ref{fig:superlattices} (${\rm R_z^i}=8.3 ^{\circ}$ and ${\rm R_{xy}}=12.6^{\circ}$), 
the CTO-like phase is strongly favored in energy.
In the CTO-like phase, as shown in Fig.~\ref{fig:superlattices} and Table~\ref{tab:bulk_value}, 
the antipolar  distortion is favored over the FE distortion due to the large trilinear
coupling term $\sim {\rm R^i_z}{\rm R}_{\rm xy}{\rm AFE_{xy}}$ in Eq.~\ref{eq:LP}.
Notably, when the amplitudes of OORs are reduced,
the BFO-like phase becomes energetically 
more stable than the CTO-like phase as shown in Fig.~\ref{fig:phase_transition}.
This indicates that the BFO-like phase can be stabilized over the CTO-like phase
when the OOR is reduced. When the above transition takes place,
the OOR around [001] will change abruptly from in-phase rotation
to out-of-phase rotation signifying a more drastic change in the topology of the oxygen octahedron network,
as guided by the yellow plane at $\Delta E =0$ in Fig.~\ref{fig:phase_transition}.
In addition to the pattern change of OOR,
the BFO-like phase is generally found to have much larger polarization
than that in the CTO-like phase as shown by the color spectrum in Fig.~\ref{fig:phase_transition}.
The much stronger FE polarization is expected, originating from the e-$R3c$ phase; it can also be
easily understood by the large four-linear coupling term
$\sim {\rm R^o_z}{\rm R_{xy}}{\rm FE_z} {\rm FE_{xy}}$, which promotes FE
in both the in-plane and out-of-plane directions.

This mechanism leads to the BFO-like phase that exists 
at the interface of the BTO/CTO superlattice.
Assuming the octahedra to be fairly rigid, the reduction of OOT is imposed by
the adjoining BTO layer, which is strongly resistant to the OOR. A direct consequence of the
stabilization of the BFO-like structure at the interface is
that the polarization of the superlattice is greatly enhanced.
For a particular choice of angles with ${\rm R_z^o}=5.7^{\circ}$ and ${\rm R_{xy}} = 6.6^{\circ}$
similar to those at the interface of the BTO/CTO superlattice,
the computed polarization of the BFO-like phase is over 54.0 $\mu C/cm^2$, which is comparable to
the polarization in the bulk e-$R3c$ CTO as shown in Table~\ref{tab:bulk_value}.
In the superlattice, the BFO-like phase is further favored by both the electric and mechanical boundary conditions
imposed by the polarization of the BTO layer according to Eq.~\ref{eq:HP}.
Under the continuous displacement field along [001] direction,
the electric boundary condition tends to polarize the CTO components with a larger ${\rm FE_z}$.
Under the tensile strain, the mechanical boundary condition effectively enhances ${\rm FE_{xy}}$.
The larger ${\rm FE_z}$ and ${\rm FE_{xy}}$ tend to further lower the energy through 
$\sim {\rm R^o_z}{\rm R_{xy}}{\rm FE_z} {\rm FE_{xy}}$ and stabilize the BFO-like phase.

\section{Inverse design of new ferroelectric materials}
It has long been recognized that oxygen octahedron rotation 
can play different roles in perovskites promoting 
 FE in BFO-like materials~\cite{NeatonBFO,Zeches-BFO}
but suppressing FE in CTO-like materials~\cite{Benedek-CTO}.
However, the results presented here suggest
that a transition between these two
phases can be achieved through interface engineering in a superlattice.
In addition to improving the fundamental understanding of these transitions,
these results suggest a new pathway to induce FE in functional oxide materials.

\begin{table}[ht]
\begin{center}
\caption{Predicted superlattices with enhanced polarizations.  
$P^{\rm {sbulk}}_{\rm A'BO_3}$($\mu c/cm^2$) 
and $P^{\rm {sbulk}}_{\rm {A''BO_3}}$($\mu c/cm^2$) denote the computed 
polarizations for strained {\rm A'BO$_3$} and {\rm A''BO$_3$} respectively.
$P_{\rm M}$($\mu c/cm^2$)  and $P_{\rm Cal.}$($\mu c/cm^2$) 
are the expected polarizations  from the {\it dielectric slab} model~\cite{Neaton_2003} 
and the computed polarizations from first-principles in $n$A'BO$_3$/$n$A"BO$_3$ respectively.
Polarization enhancement (Enh.) is calculated by ($P_{\rm Cal.}$- $P_{\rm M}$)/$P_{\rm M}$. 
Sub. denotes the proposed substrate for the epitaxial growth of superlattice.}
\begin{tabularx}{0.5\textwidth}{ l c c c c c r }
\cline{1-7}
  \hline
  \hline
{\footnotesize $n$A'BO$_3$/$n$A"BO$_3$ } 
& {\footnotesize $P^{\rm {sbulk}}_{\rm A'BO_3}$} 
& {\footnotesize $P^{\rm {sbulk}}_{\rm {A''BO_3}}$ } & $P_{\rm M}$
& $P_{\rm Cal.}$ & Enh. & Sub. \cr
  \hline
  \hline
 {\footnotesize 2BaTiO$_3$/2CaTiO$_3$} & 41.9 & 25.7 & 25.5 & 39.9 &  56\%  & {\footnotesize SrTiO$_3$}\\
 {\footnotesize 2BaTiO$_3$/2CdTiO$_3$} & 50.8 & 37.8 & 34.2 & 50.4 &  47\%  & {\footnotesize NdGaO$_3$}   \\
 {\footnotesize 2KNbO$_3$/2NaNbO$_3$}  &  36.2 & $\sim$ 0 & 19.8  & 37.5 &    89\% & {\footnotesize DyScO$_3$}\\
 {\footnotesize 2KNbO$_3$/2AgNbO$_3$}  &  36.2 & $\sim$ 0 & 22.0 & 38.2 &    73\% & {\footnotesize DyScO$_3$} \\%
     \hline
\end{tabularx}
\label{tab:example_enhancement}
\end{center}
\end{table}
The enhanced polarization observed in the BTO/CTO superlattice studied here
demonstrates this mechanism. To explore the potential of this approach, 
we predict a few more superlattices A'B$O_3$/A"B$O_3$ 
as listed in Table~\ref{tab:example_enhancement}.
Within this category of tailored materials, one of the
parent bulk A'B$O_3$ is chosen to be a ``CTO-like'' perovskite with strong oxygen octahedron rotations,
resulting in an antipolar type (CaTO$_3$ and CdTO$_3$)~\cite{Moriwake-CdTiO} or an antiferroelectric type 
(AgNbO$_3$ and NaNbO$_3$)~\cite{Fu-AgNbO,Kania-AgNbO}
ground state that is favored by the trilinear coupling term.
The other parent bulk A"B$O_3$ is chosen to have a large tolerance factor
resisting oxygen octahedron rotation and a strong FE polarization.
Similar to what we have already shown for the BTO/CTO example, 
the out-of-phase OORs around [001] are induced around the interface layers of A'B$O_3$
(see Supplementary Materials S8 for examples of 2BaTiO$_3$/2CaTiO$_3$ and 2KNbO$_3$/2AgNbO$_3$).
As a result, the overall polarizations of the superlattices are enhanced compared to the
predictions from the {\it dielectric slab} model, which is equivalent to applying the charge
continuity principle only and neglecting completely the possible interface reconstruction.

\begin{table*}[ht]
\begin{center}
\caption{Predicted new artificial ferroelectric materials based on non polar perovskites. 
(1)Mode decompositions ($\AA$) of Q$_{\rm R_{z}^{o}}$ as out-of-phase OOR around [001] , 
Q$_{\rm R_{z}^{i}}$ as in-phase OOR around [001], Q$_{\rm R_{xy}}$ as OOR around [110], 
Q$_{\rm FE_{z}}$ as polar distortion along [001], Q$_{\rm FE_{xy}}$ as polar distortion along [110],
and Q$_{\rm AFE_{xy}}$ as in-plane antipolar distortion. Their symmetries are labeled as
$M^{-}_{1}$, $M^{+}_{3}$, $M^{-}_{5}$, $\Gamma^{-}_{3}$, $\Gamma^{-}_{5}$ respectively; 
(2) OOR angles (degrees) R$_{z}^{o}$, R$_{z}^{i}$, R$_{xy}$ for 
out-of-phase OOR around [001], in-phase OOR around [001], and OOR around [110] (octahedron tilt) respectively;
(3) The polarizations ($\mu c/cm^2$) along [001] direction $P_{001}$, along [110] direction $P_{110})$, and the total polarization $P_{\rm T}$; 
(4) The polarization enhancement (Enh.); (5) Substrate for the epitaxial growth of the superlattices, KTaO$_3$(a=3.99$\AA$), MgO(a=4.21$\AA$). 
(6) Space group symmetry (Sym.) of both strained bulk and superlattice.}
\begin{tabularx}{1.00\textwidth}{ l |c c c c c c | c c c | c c c | c | c | r}
\cline{1-16}
  \hline
   \hline
 A'(A")BO$_3$ & \multicolumn{6}{c|}{Mode Decompositions($\AA$)} & \multicolumn{3}{c|}{OOR Angles($\circ$)}& 
\multicolumn{3}{c|}{Polarization($\mu c/cm^2$)} &\multirow{2}{*}{{\rm Enh.}}& \multirow{2}{*}{{\rm Substrate}} & \multirow{2}{*}{\rm Sym.}\\
{$n$A'BO$_3$/$n$A"BO$_3$}& Q$_{\rm R_{z}^{o}}$ & Q$_{\rm R_{z}^{i}}$ & Q$_{\rm R_{xy}}$& Q$_{\rm FE_{z}}$& Q$_{\rm FE_{xy}}$  
& Q$_{\rm AFE_{xy}}$   & R$_{z}^{o}$ & R$_{z}^{i}$& R$_{xy}$ & $P_{001}$ & $P_{110}$ & $P_{\rm T}$ & & & \\
  \hline
CdSnO$_3$ & 0     & 1.22  & 1.85   & 0     & 0     &0.79  & 0     &12.0  & 17.7  & 0     & 0      & 0    & -  & KTaO$_3$ &{\footnotesize Pnma}\\
BaSnO$_3$ & 0 & 0 & 0 & 0     & 0     &0     & 0     & 0     & 0    &0     & 0       & 0      & -  & KTaO$_3$ & {\footnotesize P4/mmm} \\
1CdSnO$_3$/1BaSnO$_3$ & 1.08  & 0 & 1.21& 0.20 & 0.60  & 0 & 9.4   &0 & 12.2  & 8.9 & 13.5 & 16.2 & $\infty$ & KTaO$_3$ & {\footnotesize Pc} \cr  
CdHfO$_3$ & 0     & 1.27  & 1.74   & 0     & 0     &0.69  & 0     &12.3  & 16.9   & 0     & 0     & 0      & -  &  KTaO$_3$ & {\footnotesize  Pbnm}\\
BaHfO$_3$ & 0     & 0     & 0      & 0     & 0     &0     & 0     & 0     & 0     &0     & 0      & 0      & -  & KTaO$_3$ & {\footnotesize P4/mmm} \\
1CdHfO$_3$/1BaHfO$_3$ & 1.07  & 0  & 1.17   & 0.18  & 0.64  & 0 & 9.3   &0 & 12.2  & 7.8& 10.8   & 13.3   & $\infty$ & KTaO$_3$ & {\footnotesize Pc} \\
CaZrO$_3$ & 0     & 1.02  & 1.82   & 0     & 0     &0.72  & 0     &9.2   & 16.2   &0      & 0     & 0 & - & MgO & {\footnotesize  Pcmn}\\
BaZrO$_3$ & 0     & 0     & 0      & 0     & 0     &0     & 0     & 0     & 0     &0     & 0      & 0     & -  & MgO & {\footnotesize P4/mmm}\\
1CaZrO$_3$/1BaZrO$_3$ & 0.79  & 0 & 1.22 & 0.27  & 0.71  & 0 & 6.6   &0  & 10.8 & 11.1 & 27.6  & 29.7& $\infty$  & MgO& {\footnotesize Pc} \\
1CaZrO$_3$/1BaZrO$_3$ & 0.90  & 0 & 1.16 & 0.30  & 0.67  & 0 & 8.1   &0  & 10.6 & 13.5 & 22.5  & 26.2 &  $\infty$& Relaxed & {\footnotesize Pc} \\
  \hline
\end{tabularx}
\label{tab:example_2}
\end{center}
\end{table*}

This approach to create new FE materials by interfacial control 
can also be used to create new materials even 
where the building blocks could come {\it only} from 
nonpolar perovskites. In Table~\ref{tab:example_2}, 
we list a few predicted 1A'B$O_3$/1A"B$O_3$ superlattices 
within this category.  
These interface materials also provide us a good opportunity to perform rigorous mode decompositions  
based on space group theory followed by a careful comparison 
between the interface materials and the parent bulk compounds.
The resulting mode decompositions and the local properties 
are also shown in Table~\ref{tab:example_2}. 
The A'B$O_3$ is again a  ``CTO-like'' perovskite with strong oxygen octahedron rotations.
The above property is clearly represented by the large mode amplitudes of Q$_{\rm R_{xy}}$ and Q$_{\rm R_{z}^{i}}$
which correspond to an OOR around [110] and in-phase OOR around [001] as shown in Table~\ref{tab:example_2}.
Under such a pattern of OORs, the antipolar mode Q$_{\rm AFE_{xy}}$ is favored,
and FE is strongly suppressed resulting in zero polarization along all directions. 
On the other hand, A"B$O_3$ is a strong ``cubic'' perovskite~\cite{Ahmed-BaZrO,Bouhemadou-BaHfO,Maekawa-BaSnO}
that does not display structural distortions associated with either OORs or polarization at its ground state
as shown in Table~\ref{tab:example_2}.

Strikingly, when A'B$O_3$ and A"B$O_3$ form a 1A'B$O_3$/1A"B$O_3$ superlattice,
the resulting structural distortions are significantly different from their parent bulks.
The differences come not only from the amplitudes of the modes but also from the symmetries associated with these modes.
In Table~\ref{tab:example_2}, the OORs around [110] Q$_{\rm R_{xy}}$ 
are preserved in all these superlattices but, with 
largely reduced mode amplitudes compared with those in bulk A'B$O_3$. 
In contrast, the in-phase OOR around [001] Q$_{\rm R_{z}^{i}}$ completely disappears and is 
replaced by a large mode amplitude Q$_{\rm R_{z}^{o}}$ associated with an
out-of-phase OOR around the same axis in all the predicted new materials.
As we have seen repeatedly in the previous discussions, 
such a new pattern of OOR signifies the stabilization of a ``BFO-like'' structure in all
these artificial materials. Accordingly, large polarizations develop along both
[001] and [110] directions with the generated total polarization vector roughly along the
[111] direction due to the broken symmetry in the e-$R3c$ phase.
It can be noted that the polarization of
BiFeO$_3$ is exactly along [111] direction in the $R3c$ symmetry. 
At the same time, the antipolar mode Q$_{\rm AFE_{xy}}$ is completely eliminated.
Here, we want to stress that none of the component perovskites in the predicted superlattices 
is polar either in its natural bulk or in its strained bulk formats!

\section{Outlook}
Currently, there are two widely adopted interface approaches to induce FE in oxide
superlattices, namely {\it tricolor}~\cite{Sai_PRL_2000} and {\it hybrid improper} methods~\cite{Rondinelli}.
An artificially induced broken inversion symmetry lies at the heart of both the above two
methods. In the former, the broken inversion symmetry along the out-of-plane direction is introduced by
the number of species in the superlattice; while in the latter, the broken inversion symmetry
along the in-plane direction  is facilitated by the differences in the antipolar modes
of the two perovskite materials across the interface.
However, it should be noted that the interface approach discussed here is a new route
that is conceptually different from the above. Instead of introducing
artificial inversion symmetry breaking, the ferroelectric polarization is stabilized by
favoring a ``BFO-like'' structure which is a metastable phase for many perovskite materials.
Due to the nature of the energy term that stabilizes the ``BFO-like" structure
($\sim {\rm R^o_z}{\rm R}_{\rm xy}{\rm FE_z FE_{xy}}$) , it is expected the 
switching of FE does not necessarily
require switching the directions of oxygen octahedron rotations, which
usually requires much larger energy as is implied in the {\it hybrid improper} mechanism.
Indeed, the FE polarization switching  has already  been successfully demonstrated
in 2BTO/2CTO by Lee's group~\cite{Seo-SLs}.
Based on nudged elastic band (transition state) theory ~\cite{Sheppard-NEB,Henkelmana-NEB} and single
domain assumption, the energy barrier in switching FE in 2BTO/2CTO (154 $meV$) is found to be close to that of 
the predicted materials 1CdSnO$_3$/1BaSnO$_3$ (119 $meV$) both of which are modeled in 40-atom supercells.

In conclusion, by combining HRTEM experimental and first-principles approaches,
we introduced a comprehensive interface design method to stabilize a
highly polar ``BFO-like''  metastable phase in perovskite materials.
Both the electric and mechanical boundary conditions are taken into account as well.
This scheme introduces a conceptually novel way to design artificial FE materials.
By predicting some new materials, we demonstrate this approach of exploring novel functional materials. 
For example, if the FE could be recovered
in orthogonal RFeO$_3$(R= Y,Gd, Tb Dy,Ho,Er,Tm,Yb, Lu)~\cite{Shen-YFeO,Fu-GdFeO,Zhu-LuFeO} by this approach,
the synthesis of a new family of room temperature multiferroic materials could be achieved.
Furthermore, the result of our current work indicates that, through an interface design mechanism,
short-period superlattices can have stronger FE than longer ones. This is promising for modern
device applications based on ultra-thin films.

\section {acknowledgments}

X. W. was supported as part of the Center
for the Computational Design of Functional Layered Materials,
an Energy Frontier Research Center funded by the U.S. Department
of Energy, Office of Science, Basic Energy Sciences under Award no.
DE-SC0012575. The work of K. M. R. was supported by NSF DMR-1334428.
This research used resources of the National Energy Research Scientific Computing Center (NERSC),
a DOE Office of Science User Facility supported by the Office of Science of the U.S.
Department of Energy under Contract No. DE-AC02-05CH11231.
The transmission electron microscopy was accomplished at the Electron
Microscopy Center – in the Center for Nanoscale Materials at Argonne National Laboratory,
a DOE-BES Facility supported under Contract DE-AC02-06CH11357 by UChicago Argonne, LLC.
The work at ORNL was supported by the U.S. Department of Energy, Office of Science, Basic Energy Sciences,
Materials Sciences and Engineering Division. We thank Dr. Hua Zhou for useful discussions. 
X. W. is grateful for the useful discussions with
Dr. Andrew J. Shanahan at University Medical Center of Princeton.


\end{document}